# Comparative Study of MAC Protocols for Wireless Mesh Network

Ankita Singh[1], Shiv Prakash[1], Sudhakar Singh[1*]

**Abstract**

Wireless networking is encouraged by the constant enhancement of sensors' ability and wireless communication. To provide service quality support for multimedia viz. audio and video streams, the IEEE 802.11e MAC (Media Access Control) improves basic 802.11 MAC. IEEE 802.11 standard series such as IEEE 802.11a, b, g, n, p, and ac have been promoted and specified in the current communications and connection development. Each standard has functionality that matches the kind of applications for which the standard is intended. IEEE 802.11ac has better performance with fewer interferences and achieves gigabits per second capacity transfer rates. This paper discusses the comparative examination of the IEEE 802.11a, IEEE 802.11b, IEEE 802.11g, IEEE 802.11n, IEEE 802.11p, and IEEE 802.11ac standards which increase accuracy and performance pertaining to the IEEE 802.11 standard. In this paper, we investigate the design requirements for numerous simultaneous peer-to-peer connections. Further, this study offers a systematic review and analysis of the MAC layer in WMN (Wireless Mesh Network) and also highlights their open research issues and challenges. Finally, this paper discusses various potential directions for future research in this area with an emphasis on their strengths and limitations.

*Keywords:* WLAN, IEEE Standards, Wireless Communications, Wi-Fi, IEEE 802.11, Networking.

*Corresponding author

Ankita Singh
ankita.singh87@gmail.com

Shiv Prakash
shivprakash@allduniv.ac.in

Sudhakar Singh
sudhakar@allduniv.ac.in

[1] Department of Electronics and Communication, University of Allahabad, Prayagraj, India

1. **INTRODUCTION**

Wireless mesh networks (WMNs) have grown up more attractive in the past few years, owing to their ability to provide global broadband wireless internet connection over a wide geographic region while still being relatively inexpensive to install and maintain. Many applications exist for WMNs, including commercial building automation, video surveillance, and military radar detection. In a typical WMN, routers and clients form a mesh network. As an alternative, mesh routers are high-powered stationary devices with a large number of radio ports that also function as internet gateways. Wireless mesh networks are typically used to extend single-hop infrastructure-based wireless networks. At the start, almost all wireless topologies assume omni-directional communication in WMN. This reduces network capacity by causing bad spatial reprocessing in multi-hop networks. Because of their higher spatial reuse, smart antennas can significantly increase transmission capacity. It has been recently shown that multi-beam antennas are used in wireless mesh networks.

Clients in WMNs can connect, though they are not in direct transmission range since each mesh client has in-built multi-hop communication capabilities. Mesh network is made up of the mesh routers and mesh clients. Because of the ability to route, mesh clients are able to provide better connectivity and coverage. In contrast to the ad-hoc and sensor networks, the backbone mesh enables for smooth integration of the heterogeneous wireless networks like Wi-Fi and cellular. Interoperability between laptops and mobile phones is also improved.

In recent years, MAC (Media Access Control) layer protocols for wireless mesh networks such as IEEE 802.11s, 802.15, 802.16 (WiMAX), and 802.20 have been created. As a function of this, the 802.16 standard does not allow for multi-hop client mesh topologies, distributed AAA (authentication, authorization, accounting) server authentication, and so on. Since the existing standards only allow a restricted range of WMN functions, the network's scalability and availability are constrained. As a result of the fact that these standards are derived from other wireless networks, like Mobile Ad-hoc Networks (MANETs), sensor networks, and cellular networks, their security protocols are currently in draft form. It's compatibility and integration that cause problems for WMNs when implementing these security solutions, which can leave them vulnerable. This is due to the fact that WMNs have several hops of communication and a diversified network environment, as well as dynamic network architecture and multi-channel multi-radio features. Wireless networking has experienced a lot of consideration in recent years. Mobile communication and services have become normal in the lifestyle of people with the advent of GSM (Global System for Mobile Communications). A surprising amount of growth has occurred in the past few years in the market for wireless local area network (WLAN) equipment. WLAN bandwidth can currently reach 54 Mbps, and in the future, it may exceed 100 Mbps. It is more suitable for low-cost broadband services because the spectrum is unlicensed. Due to this, many tried to extend WLAN coverage from hot spots to hot zones.

Hotspots to hot zones: Wi-Fi mesh networks provide coverage in every direction. They are mesh access points (APs), which combine access and backhaul capabilities [1]. To overcome capacity, throughput, latency, and reach issues, the WMN must be optimized. It must be carrier-class when used in an outside area. Easy implementation and inexpensiveness are the biggest advantages of the WMN. By adding or removing nodes, the mesh network's coverage area can be expanded. There is a lot of redundancy and reliability with a mesh topology. Researchers understand that WMNs have inherent functions such as organizing themselves automatically, configuring themselves, and self-healing their faults. However, IEEE 802.11's contention-based access is a limitation for real-time multimedia transmission. WLAN is governed by the IEEE 802.11 based set of standards, which was developed by the IEEE. Medium-sharing and single-hop transmission are the key features of its MAC layer protocol which is implemented using CSPAN (Cable Satellite Public Affairs Network) and CSMA/CA (Carrier Sense Multiple Access with Collision Avoidance) mechanisms whereas WMNs are characterized by multi-hop transmission. Due to medium sharing and the insufficiency of CSMA/CA, IEEE 802.11 MAC does not satisfy the needs of backhaul networking in the WMNs. The backhaul networking requires innovative MAC techniques that can assure the throughput, capacity, latency, and reach as well as Quality of Service (QoS) capability. This is a QoS-enhanced MAC protocol that can also be used in single-hop contexts (IEEE 802.11). Earlier research has shown that MAC can be enhanced. Fig. 1 shows the flowchart of link establishment and estimation of capacity for channels in MAC.

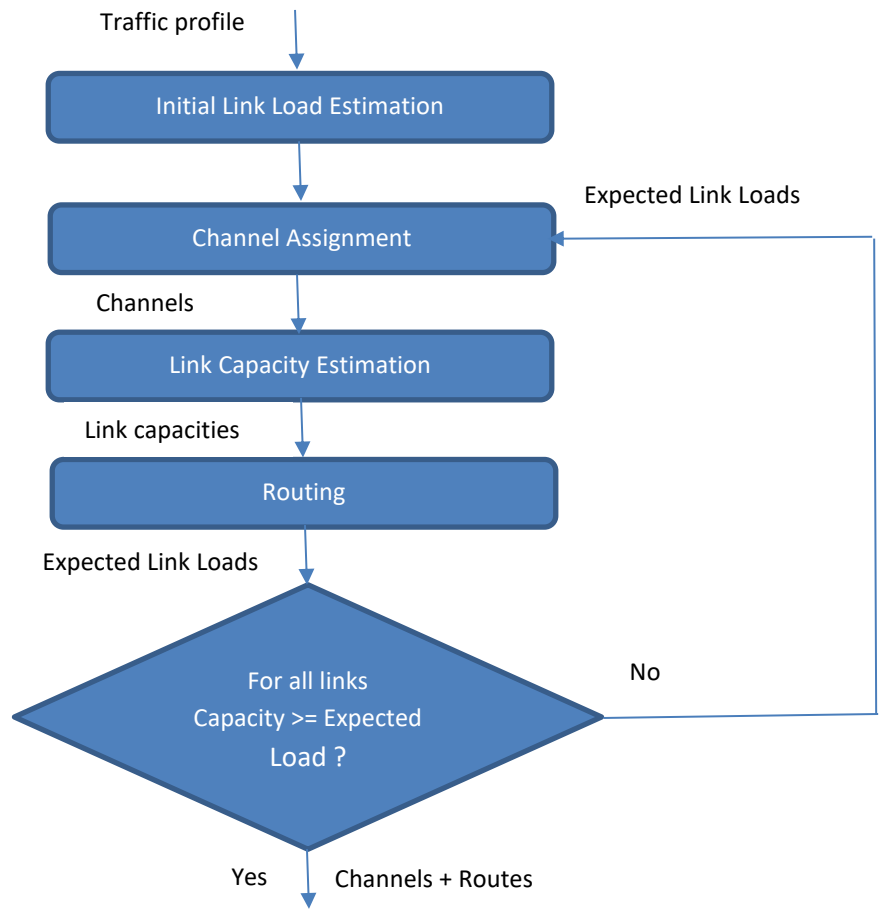

Fig. 1: The flowchart of link establishment and estimation of capacity for channels in MAC

These networks are primarily used for aircraft networking and vehicle networking, smart-cities and emergency communications as well as tactical communication due to their combination of ad-hoc network characteristics such as self-organization and self-management, self-meditated multi-hop relay, and traditional cellular network advantages for instance high-speed broadband and easy access. To meet the growing demand for data rate, fast connections between networks and systems, frequency spectrum efficiency, and anti-interference ability, as well as the accelerating development of information technology and users' constant need for larger capacity and higher data rate, a large number of in-depth experiments are required.

This paper addresses and presents a detailed comparison of common Wireless LAN (WLAN) technologies such as IEEE 802.11a, IEEE 802.11b, IEEE 802.11g, IEEE 802.11n, IEEE 802.11p, and IEEE 802.11ac which helps to understand the advantages and disadvantages, and applicability of each standard in various situations.

The rest of this paper is structured as follows. Section 2 deliberates the wireless mesh network. Section 3 describes medium access control tools. Section 4 discusses the literature review of the relevant studies on MAC protocols for WMNs. Section 5 covers the comparative analysis of IEEE 802.11 protocols. Section 6 confers the summarized observations of the study, and finally, Section 7 discourse the conclusion of the paper.

## 2. WIRELESS MESH NETWORK

The use of mobile communication devices, for instance, tablets and smartphone devices have increased dramatically during the past few years. While using these mobile devices, mobile consumers traffic more data. According to the report [2] of 2015, as indicated in Table 1, it is estimated that there will be several mobile device subscription prospects between 2014 and 2020.

Table 1 shows a huge number of mobile devices and the fast growth of customers. It should also be pointed out that growth in mobile PCs, routers, and tablets is significantly less as compared to high-speed mobile devices and smartphones. In the communication sector, new connectivity technologies have emerged to facilitate this expansion. Wi-Max, Wi-Fi, LTE, 4G, and 3G are available. These technologies follow different standards on the construction and communication of systems. WLAN with IEEE 802.11 standards are used by mobile broadband devices whereas LTE, 4G, and 3G technologies are used by cellphones respectively.

Table 1: Outlook for mobile subscription (in million) as set out in [2] [3]

| Subscriptions | In year 2014 | In year 2020 |
| --- | --- | --- |
| Total mobile | 7100 | 9200 |
| Mobile broadband | 2900 | 7700 |
| Smartphones | 2600 | 6100 |
| Mobile PCs, Tablets, and Routers | 250 | 400 |

In recent years, wireless networking has been considered as an emerging area for research and the industry. Local Area Network (LAN) / Metropolitan Area Network (MAN), these two specifications include features of MAC Layer and Physical Layer (PHY) which are the parts. Several questions were also resolved through the investigation of MAC operations in IEEE 802.16 and IEEE 802.11. These findings are calculated with different measures including performance, equity, latency, and use and drop rate.

The first WLAN norm was adopted by IEEE in 1997. A networked LAN links customers (computers, laptops, phones) with access points to offices and networked public areas. Wireless LAN named IEEE 802.11 offers networking connectivity. Often clients are attached to the WLAN network's central access point. It can be used in two ways, direct or grouped, the relationship between customers and points. Communication to or from a client also takes place via a single connection point. Contacts between two customers take place without the need for a connection point in clustered mode. The IEEE 802.11 contains physical and internet communication TCP/IP layers that combine wireless networking within the ranges of 2.4, 3.6, 5, and 60 GHz frequencies. A TCP/IP protocol with five layers includes physical addresses, logical addresses, network addresses, and single addresses.

The 802.11 Protocol defines the following two modes: the PCF (Point Coordination Function) and the DCF (Distributed Coordination Function). The novel Hybrid Connectivity Feature (HCF) is now included in the 802.11e protocol, which offers several implementations to meet customers' demand for real-time applications. The HCF is split into two categories: EDCA (Enhanced Distributed Channel Access) and HCF Controlled Access (HCCA). HCF consists of two modes of operation. The key connectivity methods of the Access Point (AP) node are PCF and HCCA. In compliance with a pre-set time, the AP uses polling to award channel access privileges. The downside of providing a control node and adding an overhead polling message is usually delivered by reduced physical costs for all operations. In comparison, the DCF and EDCA connection-based access mechanisms are distributed which define the right of access on the wireless platform by various local containment parameters used by any device. Enlarging DCF, EDCA provides a range of QoS (quality of service) indicators, including priority levels and deadlines. The importance of transmitting IEEE 802.11, particularly DCF, the basic business of the MAC Protocol as out in all IEEE 802.11 standards, comprising IEEE 8002.11e, is significant. Stations provide a random back-off monitoring method for the channel using the IEEE 802.11 DCF. Unused slots and collisions are also acted as overhead that decreases DCF efficiency. This failure increases due to higher loads and network sizes and secret terminals. A mixture of IEEE 802.11e and IEEE 802.11 standards can also be implemented into wireless networks. As a result of the deference between EDCA and conventional DCF, the reliability of these networks was also of interest.

Wi-Fi-MESH Wireless network is used for aircraft networking, vehicle networking, smart cities, emergency, and tactical connectivity in the fields of war. It also incorporates characteristics of an ad-hoc network that is self-organized, self-managed with multi-hop relays, and the advantages of broadband are high speed, fast access, and conventional connectivity. Nevertheless, vast quantities of in-depth testing are expected, for increased data rate demand, fast communications between networks and devices, and frequency spectrum reliability and anti-interference capability, for accelerated information technology advancements and for continuously increasing usage ability and data speeds. Two main innovations in Wi-Fi-MESH's wireless network are media access control (MAC) and network routing technology. The general architecture of WMN is shown in Fig. 2.

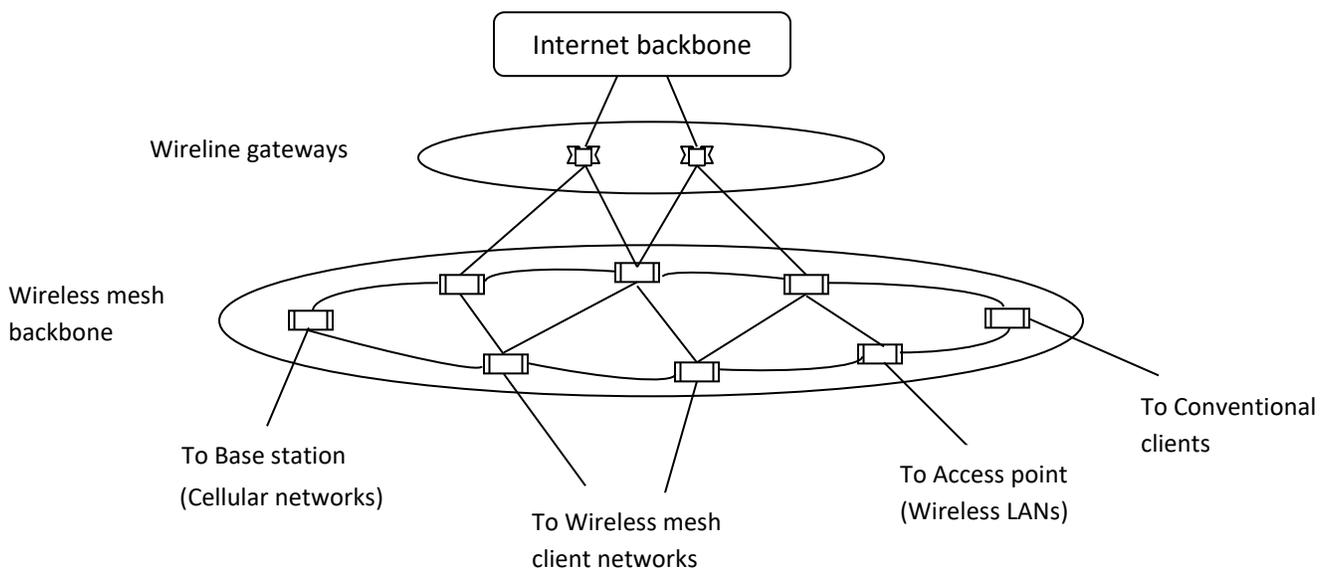

Fig. 2: Wireless mesh network [4]

### 3. MEDIUM ACCESS CONTROL (MAC)

Links must be correctly constructed in order to make use of multi-beam smart antennas. Protocols described by IEEE 802.* specifications are used in WLANs and ad-hoc networks. For example, the MAC protocol is designed to establish rules for the efficient and equitable sharing of the common wireless channel [5], [6]. Contention-free and contention-based are the two basic types of MAC protocols for the wireless networks [7]. Contention-free MAC is based upon the controlled access, where channels are allocated to each node according to a timetable. Random access is used to implement the contention-based MAC protocol, in which nodes compete for access to shared media. Disagreements are resolved by using the distributed conflict resolution procedure.

When it comes to wireless communication, IEEE 802.11, the de facto wireless network media access control standard, is built for omnidirectional antennas. DCF is a compulsory channel access function in the IEEE 802.11 standard, while PCF is an optional channel access function. As a result, PCF is more centralized while DCF is more dispersed. The IEEE 802.11e standard, an improvement of 802.11, alters the MAC layer so that it offers a set of the quality of service (QoS) upgrades for WLAN applications [8]. Voice over Wireless LAN (Vo-WLAN) and streaming multimedia rely on this protocol.

## 3.1 The General Objectives of MAC Protocols in WMN

Unlike Ethernet (IEEE 802.3), designing a MAC protocol for wireless LANs is fraught with difficulties due to the fluctuations found in the radio channel. To provide wireless devices with access to WLAN radio channels, a controlled and distributed infrastructure is necessary. Because of CSMA, wireless networks' most widely used medium access protocol, MAC protocols were devised to take into consideration various elements of the shared radio channel. As part of the process of creating a half-duplex and full-duplex MAC protocol, numerous frequent issues must be addressed. In addition to these complex issues, the protocol must solve a few fundamental issues also known as performance parameters for IEEE802.11 networks as discussed below.

### 3.1.1 Throughput

A radio link's data transmission capacity is measured as a percentage of the radio link's capacity. Numerous factors influence it. These include MAC protocol efficiency, overheads, multiple access interference, and channel transmission impairments to name a few. Comparatively speaking, a Flow Driven Media Access Control (FD-MAC) protocol is expected to treble the throughput of standard half-duplex MAC protocols. An FD-MAC must be able to overcome the issues that come from the simultaneous broadcast and receiving of radio signals in the same frequency band [9].

The maximum throughput achieved by an application is expressed by the Eq. (1) when there is no fragmentation involved at the lower layers.

$$TMT_{APP} = \frac{\beta}{\alpha+\beta} \times TMT_{802.11}(bps) \qquad (1)$$

where, $TMT_{APP}$ is the Theoretical Maximum Throughput of the application layer, while α and β denote the total overhead above MAC layer and the size of application datagram respectively. $TMT_{802.11}$ is the theoratical maximum throughput of 802.11 MAC layer [10].

### 3.1.2 Delay

The MAC layer delay is the total of time a frame must wait at the sender node. When a frame is placed in the sender node's MAC queue, and when it is removed from this queue, the difference between the two times is measured. When used in time-critical applications, FD-MAC can lower the average delay of each frame to increase throughput [9].

According to Chatzimisios et al. [11], when two or more competing stations select the same backoff slot for transmission, collisions occur. By assuming a negligible frame drop probability, the authors defined the total average frame delay E[D] as given by Eq. (2) [11].

$$E[D] = E[X] \cdot \sigma + E[N_f] \cdot (P_s \cdot T_s + (1 - P_s) \cdot T_c) + T_g \qquad (2)$$

where, E[Nf] is the mean of the number of times the counter of a station freezes because of other transmissions and E[X] is the time required for empty slots before successful transmission [12].

### 3.1.3 Fairness

It's important to ensure that each node has the same amount of time to access the media when using MAC. When using wireless technology, a distributed MAC protocol is most desirable. To avoid favoring one node over another, the medium access mechanism need to be included in the distributed MAC protocol architecture. Other types of data traffic, such as those originating from time-critical applications, may be given more priority. The same priority traffic class must have equal access to the medium in order to support application fairness. Fairness should be considered while assigning radio channels to full-duplex nodes [9].

The Jain's fairness index [13] is a broadly used metric for the evaluation of the fairness of resource allocation amongst traffic flows. Eq. (3) defines the Jain's fairness index.

$$J_I = \frac{(\sum_{i=1}^{S} T_i)^2}{S \times \sum_{i=1}^{S} T_i^2} \quad (3)$$

where, $S$ is the number of stations and $T_i$ is the throughput of $i^{th}$ station. Alternatively, it can be also expressed as Eq. (4) and Eq. (5).

$$J_I = square\ of\ the\ sum\ of\ all\ throughputs/(S \times sum\ of\ squares\ of\ all\ throughputs) \quad (4)$$

$$J_I = square\ of\ average\ throughput/average\ of\ squared\ throughput \quad (5)$$

### 3.1.4 Energy Consumption

The recent trends in WMNs research are being based on optimizing multi-hop protocols to increase the scalability and channel capacity. For a large time span, not much percentage of the capacity of these devices is actually used and further goes to high energy waste. Hence, it is possible to achieve the maximum power saving by turning off unnecessary network devices. Energy efficiency is a very critical factor to work on, it can be measured by Eq. (6) [14].

$$Energy\ consumption\ (E) = (T_xPowerCoeff \times t_xPower + T_xPowerOffset)\ t_xDuration \quad (6)$$

where $T_xPowerCoeff$ denotes the Transmit Power Coefficient, $T_xPowerOffset$ is the Transmit Power offset, $t_xPower$ is the transmitted power which is proportional to the distance the signal is supposed to travel, and $t_xDuration$ is the time taken for transmission of signals to reach the receiver. The values of $T_xPowerOffset$ and $T_xPowerCoeff$ are statically defined based on the WLAN specifications which are assigned the values of 16/sec, and 90mW respectively [9].

### 3.1.5 Reliability

Reliability is demarcated as the probability that a system, service, or product will perform its envisioned function adequately for a specified period of time. Reliability is often calculated as the Packet Delivery Ratio (PDR) by Eq. (7), very widely used in a number of literature [9].

$$PDR = Total\ number\ of\ packets\ received\ /\ Total\ number\ of\ packets\ sent \quad (7)$$

### 3.1.6 Latency

Latency means the time taken in capturing, transmitting, processing a data packet through multiple devices, and decoded after receiving it at the destination. It is also defined as the time gap between when a packet is sent and the time at which it is received [9]. In other words, if $T_S$ be the time a sender node sends the packet and $T_R$ be the time when a receiver node receives the packet, then it is defined by Eq. (8). Note that if a packet of data is lost then the latency will be skewed.

$$Latency = T_R - T_S \quad (8)$$

## 4. LITERATURE REVIEW

Low-cost wireless access is now dominated by IEEE 802.11x MAC layer and physical layer (PHY) technologies [15]. Facilitating high-throughput applications, for example, remote desktop and video streaming with high-definition, on the other hand, remains a challenge when there are multiple concurrent users [16]–[21]. It is for this reason that task group *n* came up with 802.11n [22], which uses several antenna systems to improve the rate of data in a single link while enabling restriction for a single user at a time. Antenna systems that permit multiple simultaneous communications have been the subject of numerous theoretical [23]–[27] and numerical/experimental investigations [5]–[7], [15]–[20], [28]–[34]. A single source transmits numerous packets simultaneously to various destinations, whereas a single destination receives multiple packets simultaneously. Concurrent communication between distinct node pairs is referred to as concurrent communication message-passing concurrency (MPC). All transmissions must begin at the same time in MPT/MPR (Multi-Packet Transmission/Reception) scenarios, hence synchronization is required. It is also possible that there is asynchronous access in MPC conditions since the transmission times of each source can be set independently [26], [27]. For downlink processes among the access point and the various users, MPT is the first of these three strategies to be implemented and its functionalities, which rely on multiuser multiple input multiple output (MU-MIMO) technique, are supported in the Long-Term Evolution Release $8^{th}$ and confirmed in the standard 802.11ac. When used in centralized contexts, where synchronization is assured by the access point [35]–[38], MPT/MPR are commonly recognized as acceptable [39]. Asynchronous access instead can deliver a high output than synchronous access, as seen in [27]. As a result of this, smart antenna systems can't be fully utilized in a distributed environment due to the possibility of change in the distribution of active transmitters while the source to destination communication is in place. To compensate for the lack of centralized control, several MPC extensions for 802.11 DCF have been proposed [5]–[7], [15]–[20]. It is vital to take into account three important properties of these MPC theories [18]–[20]. Even minor modifications to the access rules can threaten 802.11 backward compatibilities, preventing nodes employing conventional access mechanisms from connecting. On the other hand, there are nodes with varying antenna systems in the scenario, and it's important to prevent the more advanced nodes from monopolizing all resources. Last but not least, the access scheme's ability to communicate position and traffic related information to each node is critical to achieving as close as possible to central network performance [40]. Two new MAC protocols have been designed based on these standards. According to [41], the first protocol, named threshold access MPC (TAMPC), is based on an access policy determined by a sustainable load threshold that is reliant on single-node antenna capabilities. It is based on a metric called sustainable code rate which is determined using an accurate calculation of the sustained information rate (SIR) and specified channel code rate. They are tested on different types of network scenarios by using a realistic simulation environment and distinct performance metrics. Wang and Qin [42] and Tian et al. [43] have carried out an extensive survey on MAC Protocols for WMNs with Multi-beam Antenna techniques and Wi-Fi HaLow for Internet of Things (IoT). Ahmed and Misra [44] examines the feasibility and efficiency of a relay-based IEEE 802.11ah channel access strategy for multi-hop and large-scale IoT networks.

According to the multi-hop relay security standard of IEEE 802.16j-2009 [45], there are three levels of hierarchies, the Mobile Station (MS), the Base Station (BS), and the Relay Station (RS). A Security Association (SA) was first created by the Red-Security and the British Security Service (BSS). Mobile Station sends an authentication request to RS if the Burst Scheduling Problem (BSP) isn't in the mood. If the Base Station (BS) is not in the range, the RS passes the message on to the next subordinate RS, and so on. This is followed by RS creating SA between MS and the subordinate RS repeating this process until the base station receives the request. For example, if an MS sends an incorrect request, the request must be transmitted via intermediate RS until the master BS identifies the fraud id Message-Message (MM), which implies that the entire RS and SA association process is a waste of time and resources. A single base station makes collusion attacks more severe in the IEEE 802.16j. But heterogeneous networks cannot be communicated using this standard. Ad-hoc and sensor networks cannot use this security standard due to its considerable communication and processing overhead.

The Challenging issue in WMN is the enhancement of throughput which can be suitably solved by MACA-P (Medium Access via Collision Avoidance with Enhanced Parallelism) [46]. To overcome the problem of concealed and exposed nodes, the MACA-P concept [46] was created. To solve the problems of hidden and exposed nodes, as well as increase flexibility and scalability, the employment of the smart and directional antennas is crucial. According to the IEEE 802.11 MAC protocol specification, all the other nodes in the zone must remain silent when two nodes communicate over a channel using omni-directional antennae. Using directional antennas, there have been suggestions for overcoming this problem and enhancing simultaneous communication between nodes [47].

Both MIMO and Multi-Radio Unification (MRU) [28] have become important and recommended WLAN technologies for multi-channels. In addition, Maheshwari et al. [30] investigated the use of Code Division Multiple Access (CDMA) for the wireless ad-hoc networks. Nearly all of these, however, are still plagued by problems. We must deal with concerns such as directional antenna deafness TDMA's (Time Division Multiple Access) temporal synchronization, and CDMA's code management [31], [32]. Due to this, a number of changes should be made to the protocols so that they can perform better inside their intended network environments. As it can be seen that each of the works listed above has its own set of problems and rewards.

Banerji et al. [48] proposed that it should serve as a series of instructions for applying wireless communications at frequencies of 2.4, 3.6, 5 among 60 GHz. In schools, colleges, offices, homes such as public or private networks, wireless internet, mostly used in shorter-haul communications has become more widespread. WLAN has a lot of versatility but has a range of problems as well. A loss of productivity as the number of stations grows is one of the main downsides. In [49], have evaluated MANET (Mobile ad-hoc network) output in compliance with the IEEE 802.11a and 802.11b standards and a modified connection state routing protocol. The author compared the performance of the small and wide devices with different criteria of service quality, such as the network initialization, latency, copying, hello transfer, routings, and data obtained in the two different IEEE 802.11a and 802.11b protocols. The efficiency of the OPNET simulator is calculated and 50 nodes are required for two scenarios in one situation.

The AODV (Ad-hoc On-demand Distance Vector Routing) and the DSR (Dynamic Source Routing) protocols were tested by Dharamvir et al. [50]. The next standard for connectivity stability is IEEE 802.11e. The QualNet simulator IEEE 802.11e output should be assessed for the measurement of the device type such as normal jitter, cruise speed, end-to-end retardation, intake, and idle mode.

In [51], the IEEE 802.11a and IEEE 802.11g AODV, OLSR (Optimized Link-State Routing Protocol), and GRP (Geographic Routing Protocol) routing algorithms are widely used. The performance is calculated using various physical features and numbers of nodes depending on network load, latency, propagation activities, and delays in media access. The findings show that OLSR in each case performs best in terms of time and performance of media access. In contrast with AODV and OLSR protocols for 75 and 150 nodes with IEEE 802.11a, GRP is showing improved performance in retransmission. In all three protocols, the network load is nearly the same, with nodes being 75 for 802.11a and 802.11g. GRP provides a low network load for OLSR for 150 nodes.

A WMN is made up of a dynamic group of Mesh Access Points. It is comparable to an ad-hoc network, except the topology is static and made up of Mesh Access Points. The Mesh Access Point provides more features than an Access Point in terms of providing real-time multimedia transportation. The IEEE 802.11 MAC protocol, CSMA/CA, was created for single-hop WLAN environments rather than multi-hop wireless networks. Two major issues are highlighted as follows.

a) *Backhaul Networking:* The following issues must be addressed in WMN's backhaul networking: capacity, throughput, latency, and reachability. But the analysis in the literature [52], [53] shows that CSMA/CA can not meet the criteria of backhaul networking. Xu and Saadawi [52] discovered that CSMA/CA do not perform well in a wireless multi-hop environment. TCP is used as the transport layer protocol to estimate the throughput behaviour.

b) *Multimedia Transportation:* The IEEE 802.11e is a QoS protocol for wireless networks. Is it capable of performing well and ensuring QoS for multimedia traffic in WMNs? The authors in [54], [55] demonstrated that the performance of ad-hoc networks using Enhanced Distributed Coordination Function (EDCF) is suboptimal because the EDCF parameters cannot be adjusted to the conditions of the network. The collision rate enhances faster in ad-hoc networks once the frequency of contentions to admittance of the shared medium is high. It considerably affects the throughput, the latency, and therefore reduces the show of delay-bounded traffic.

## 5. COMPARATIVE ANALYSIS OF IEEE 802.11 STANDARDS

This section compare and contrast the six IEEE 802.11 standards namely 802.11 a/b/g/n/p/ac. A comparative analysis of these protocols is also carried out with respect to the quantative and qualitative parameters.

1) **802.11a:** This protocol [56] was introduced at the same time as protocol 802.11b in 1999 [57]. It offers a more powerful 54 Mbps with 5 GHz high-speed data transmission system. This high frequency reduces the bandwidth of the network from 802.11b. 802.11a, however, faces more difficulty for entry to the walls and other barriers. This protocol uses the OFDM (Orthogonal Frequency Division Multiplexing) encoding algorithm rather than an FHSS (Frequency-Hopping Spread Spectrum) or DSSS (Direct Sequence Spread Spectrum) orthogonal frequency division. In the unified band, 802.11a is smaller but its production was appealing, while the equipment required for achieving this performance was comparatively costly.

2) **802.11b:** It follows the basic streamlining technique which was laid down in 1999 [57]. It is 100 meters in range, but it is not at its peak data volume. For the transmission of data in 802.11b, the CSMA/CA (Carrier Sense Multiple Access with Collision Avoidance) techniques are used. The highest transmission rate of data is 11 Mbps in the 2.4 GHz band. A higher data rate of 5.5 Mbps will be obtained by the system and 2 and 1 Mbps if the signal is decreased, or the interruption frequency is raised. The method here used is called adaptive rate selection (ARS).

3) **802.11g:** The third cellular LAN modulation standard launched in 2003 is 802.11g [58]. The frequency spectrum resides in band 2.4 GHz and the data rate is high, 54 Mbps, using the CSMA /CA protocol for the transmission process. The 802.11g modulation schemes are orthogonal division multiplexing. Four separate physical layers, of which ERPs (Extended Rate PHY) are defined technically as extended scale, are used to achieve maximum power. These occur together during the exchange of frames, such that one of the four is chosen by the sender, as it is acknowledged at either end of the chain. The four-layer choices are listed for the 802.11g specifications: ERP-DSSS-CCK (Extended Rate Physical Direct Sequence Spread Spectrum Complementary Code Keying), ERP-OFDM (Extended Rate Physical Orthogonal Frequency Division Multiplexing), ERP-DSSS / PBCC (Packet Binary Convolution Code), DSSS-OFDM (Direct Sequence Spread Spectrum Orthogonal Frequency Division Multiplexing).

4) **802.11n:** It is published in 2009 [21], attempts to enhance current specifications by upgrading the signal processing system and expanding the MAC network using a range of antennas. IEEE 802.11n also added two new approaches, frame consolidation, and block recognition to boost the performance of the MAC network [59].

   Fig. 3 shows the channel breakdown in the 802.11n mixed mode. Of the total, there are 14 channels which are designated in the 2.4 GHz range and spaced 5 MHz apart from the exception of 12 MHz spacing before each channel. It is not feasible to assure OFDM operation which affects the number of probable non-overlapping channels depending on multiple radio operations.

5) **802.11p:** In 2010, IEEE 802.11p introduced a new Wi-Fi Networking (WAVE) Protocol 802.11p [27]. The term 802.11p is for the dedicated short-range communications (DSRC) which aim to develop the IT system (ITS). ITS vehicle services (V2V) infrastructure vehicles (V2I), as well as pedestration vehicles, have recently been the most commonly used vector in the auto industry to share data between high-speed vehicles within the 5.9 GHz ITS-licensed range.

6) **802.11ac:** This standard was adopted in 2014 [60] and ensures the maximum performance in the 5 GHz channel of 1000 Gbps. Considerations include larger RF (radio frequency) channel diameter (up to 160 MHz), however, in the case of 802.11n, this bandwidth would vary between 40 MHz and 80 MHz, a further MIMO space stream up to 80 although, in the case of 802.11n, only four space sources would occur.

**Non-overlapping channels for 2.4 GHz WLAN**

**802.11b (DSSS) channel width 22 MHz**

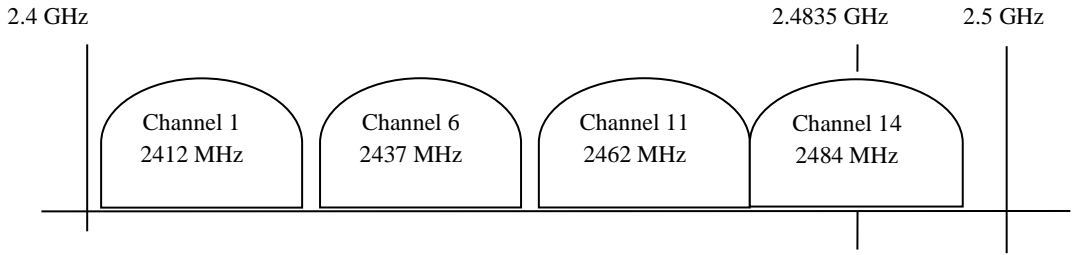

**802.11g/n (OFDM) 20 MHz channel and bandwidth 18.25 MHz used by subcarriers**

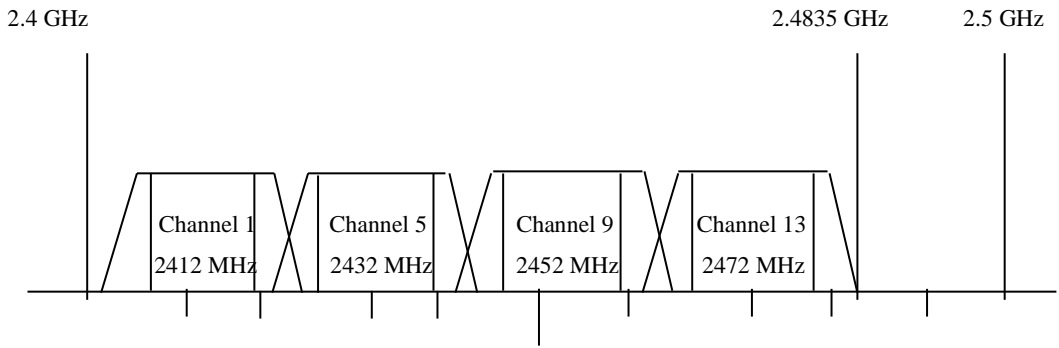

**802.11n (OFDM) 40 MHz channel and bandwidth 33.75 MHz used by subcarriers**

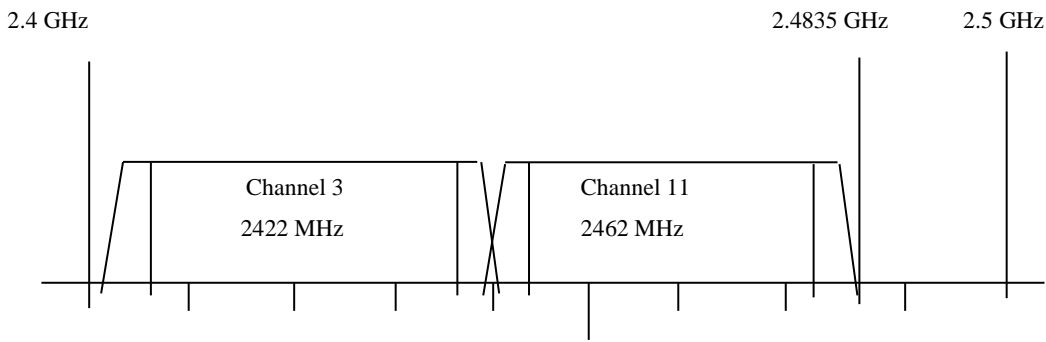

Fig. 3: The channel breakdown in the 802.11n mixed mode [61]

Fig. 4 shows the ability of an IEEE 802.11ac device which detects the transmission on its primary channel as well as on secondary channel also. The clear channel assessment (CCA) capabilities of the secondary channels were also restricted, hence deploying two 802.11n networks that overlain required in reality that the primary channels be alike identical. It has adequately good CCA capabilities on the secondary channel that two networks can easily be deployed without being overlapped, leading to performance gain for the entire network since a much bigger portion of transmissions can be carried out in parallel. Such single subtle specification enables a wide-ranging deployment option for the 802.11ac networks.

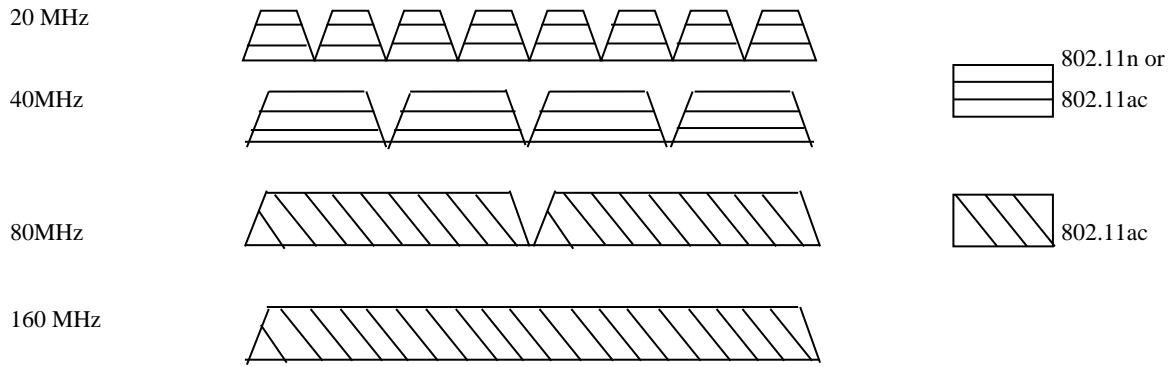

Fig. 4: Channel scaling capability from 20 MHz to 160 MHz at 802.11ac [62]

The IEEE 802.11 and its extensions are represented in Table 2. Table 2 shows a very familiar comparison of seven IEEE 802.11 standards in terms of maximum data rate, RF band, used technique, channel bandwidth, and range [15][21], [56]–[58][63][64][65]. In Fig. 5, the performance of different IEEE standards are visualized with respect to the maximum data rates in Mbps and range in meters.

Table 2: IEEE Standards Comparison on Different Parameters [65]

| 802.11 Standards | Year | Used Technique | RF Band (in GHz) | Channel Bandwidth (in MHz) | Maximum Data Rate (in Mbps) | Range (in meters) |
|---|---|---|---|---|---|---|
| 802.11 | 1997 | DSSS | 2.4 | 20 | 2 | 100 |
| 802.11a | 1999 | OFDM | 5.0 | 20 | 54 | 80 |
| 802.11b | 1999 | DSSS | 2.4 | 20 | 11 | 100 |
| 802.11g | 2003 | DSSS, OFDM | 2.4 | 20 or 40 | 54 | 100 |
| 802.11n | 2009 | OFDM | 2.4 and 5.0 | 20 or 40 | 600 | 140 |
| 802.11p | 2010 | MIMO, Frame Aggregation | 5.9 | 10 | 27 | 1000 |
| 802.11ac | 2014 | OFDM | 5.0 | 60 or 80 | 1000 | 160 |

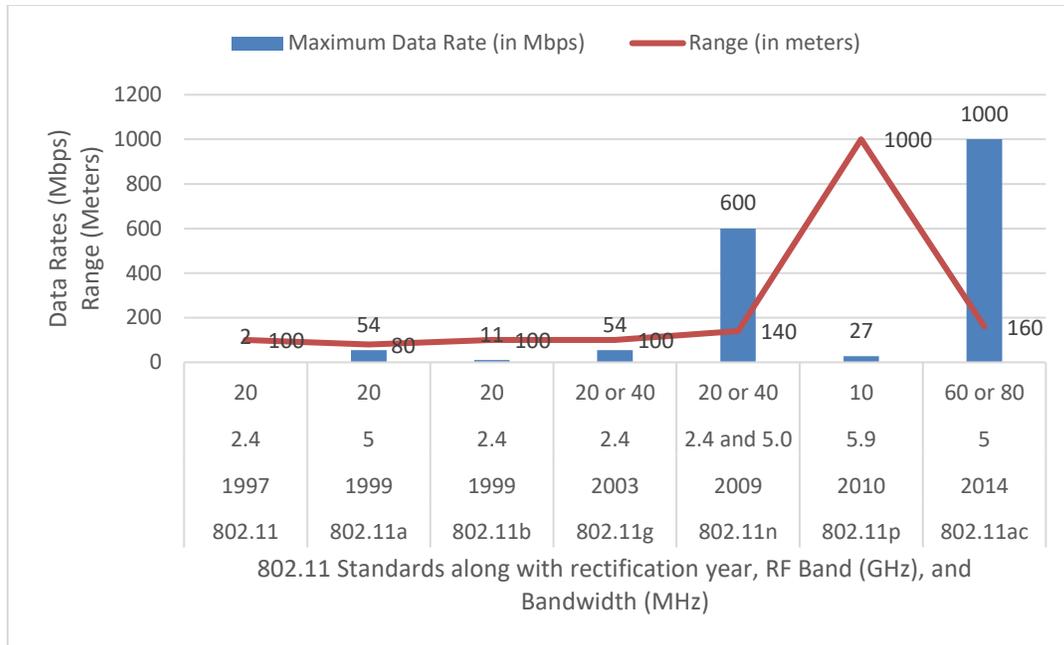

Fig. 5: Performance Comparison of different IEEE standards with respect to Data Rate and Range

IEEE 802.11 standards are experiencing low-cost equipment for end-user and high-speed data rates due to wider acceptance. Firstly, IEEE 802.11 standard has data rates of 1 and 2 Mbps based on DSSS, Frequency Hopping Spread Spectrum (FHSS), and Infrared (IR) techniques for three different physical layers. CSMA/CA is the basis of MAC protocol in 802.11 WLAN, which allows a single user at one time to utilize the radio channel. Subsequently, various new specifications for the PHY layer were incorporated though the MAC specification continued mainly unaffected.

### 5.1 Qualitative Analysis of IEEE 802.11 Standards

A qualitative analysis of the standards of IEEE 802.11a, 802.11b, 802.11g, 802.11n, 802.11p, and 802.11ac on some other popular aspects like beam formation, range, power, interference, and efficiency is given in Table 3 which is reproduced from [3][66]. These qualitative features are defined as follows.

1) **Beamforming:** Beamforming is a signal processing technique that calculates how data can be transmitted or obtained in a directional beam. This functionality is presented in both IEEE 802.11n and ac, while other specifications have no stratifying capability.

2) **Coverage and Capacity:** IEEE 802.11ac has a wide coverage against other standards, as stated earlier. It has improved link speeds with MU-MIMO and multi-space interfaces [67].

3) **Quality and Interference:** The IEEE 802.11ac operates at a 5 GHz frequency and is less vulnerable when used at 2.4 GHz than IEEE 802.11b. For IEEE 802.11a, n these standards are more likely to be jeopardized when operating with 2.4 GHz.

Table 3: Qualitative analysis of the standards [3]

| Qualitative Features | 802.11a | 802.11b | 802.11g | 802.11n | 802.11p | 802.11ac |
|---|---|---|---|---|---|---|
| Beamforming | No | No | No | Yes | No | Yes |
| Coverage | Low | Low | Low | Low | High | High |
| Capacity | Low | Low | Low | Low | Low | High |
| Interference | More on 2.4 GHz Less on 5 GHz | More | More | More on 2.4 GHz Less on 5 GHz | Less | Less |
| Quality | Low | Low | Low | Low | Low | High |

**5.2 Challenges Associated with Peer-to-Peer Connections**

All of the systems either computers or workstations that are linked with a wireless network interface card are able to communicate with one another through the use of radio waves when operating in ad-hoc mode, which is also called Independent Basic Service Set (IBSS) or peer-to-peer mode. No access point is required for this connectivity. When it comes to swiftly installing a wireless network in a conference room or centre, or any other location where appropriate wired infrastructure does not exist, ad-hoc mode is a simple option [68].

Peer-to-peer (P2P) networks derive their infrastructure from IP routing and are primarily intended for wired networks. The implementation of current peer-to-peer protocols in wireless multi-hop networks necessitates modification. However, the majority of DHT-based protocols like as Chord, Astroy, Tapestry, and others do not necessitate any modifications and may be readily implemented at the network layer in a wireless multi-hop network. Peer-to-peer networks have various issues, including the dissemination of malicious content, the practice of free riding, the occurrence of whitewashing, inadequate search scalability, and the absence of a strong trust model. In a wireless multi-hop network, peer-to-peer applications face various challenges, including but not limited to bandwidth constraints, restricted power backup, overlay maintenance caused by node churn and mobility, routing-scope, mobility management, and security concerns. The node churning and mobilization provide challenges in terms of topology maintenance and resource searches [69].

The implementation of peer-to-peer protocols in the wireless multi-hop networks presents additional issues, particularly in terms of managing the mobility of mobile nodes, addressing intermittent connections, and managing restricted bandwidth. The challenges of mobility management, short power backup, and restricted gearbox range are of significant importance and warrant more investigation in future research endeavours [69].

## 6. DISCUSSIONS

IEEE 802.11's main feature is the capacity to transport data at a fast speed. This enables the connection to a remote host to be desirable and offers better bandwidth and cheaper recurrent costs compared to prior systems. Several versions of the standard can also be available to assist users to choose a technology fit for their requirements. It offers a versatile technique to achieve data transmission across different devices.

This paper discusses how IEEE 802.11a, b, g, n, p, and ac standards, increase accuracy and performance relative to IEEE 802.11. The newest platform introduced in December 2013 is IEEE 802.11ac for hitting Gigabits per second capacities and transfer rates. As the standard operates in 5 GHz, IEEE 802.11ac reveals greater performance with less interference. It provides a diverse spectrum of 20, 40, 80, and 160 MHz widths. Up to 8 space sources are provided and antenna technologies are used by OFDM and MU-MIMO [70].

*(i) IEEE 802.11a*
It offers the higher frequencies and speed. Since many users and equipment utilize the 2.4 GHz frequency spectrum significantly, it provides less interference with the move to the 5 GHz range. With the 5 GHz carrier frequency, 802.11a is restricted to nearly a sightline, with more access points to be used. It also entails a significant reduction in signal penetration via various barriers compared to IEEE 802.11b. This standard is also less desirable for higher prices and a shorter range. Further, it is not compatible with 802.11b, one of the main drawbacks.

*(ii) IEEE 802.11b*

For a good signal range, the IEEE 802.11b standard offered a low-cost substitute. However, compared to other standards, the offered speed is slower. In addition, the 2.4 GHz frequency band is severely overcrowded and 802.11b is affected by networking devices like Bluetooth, microwave ovens, and mobile phones.

*(iii) IEEE 802.11g*

The IEEE 802.11g may be considered as an 802.11b superset that has a higher OFDM transmission performance, and support all features and backward compatibilities. It offers a comparatively better portfolio than the 802.11b standard, even when operating in DSSS mode. Since 802.11g relies significantly on the technology of 2.4 GHz, it has an additional pro from substantially reduced costs and savings due to the growing number of devices based on 802.11b. As a consequence, 802.11g equals the cost of 802.11b equipment.

*(iv) IEEE 802.11n*

Wireless LAN capacity and the overall performance of all devices can be boosted by 802.11n considerably. In a 5 GHz frequency band, there are significantly more channels than in a 2.4 GHz. The 20 MHz channel works on a 2.4-GHz frequency band and a 40-MHz channel with a greater data rate and increased efficiency, with a single 802.11n access point. One of the key concerns of this standard may be the usage of the 40 MHz 802.11n operating modes in 2.4 GHz bands since the interference of a single 40 MHz transmitter may affect a considerable part of the band. In fact, the connection of a 40-MHz channel with the original 20-MHz channel is necessary in order to release any other legacy device that operates within the same channel. There is a substantially limited opportunity for 40 MHz activities in this frequency. Even removing all outdated 802.11b and 802.11g devices from the spectrum, it is difficult to use access points on the 40 MHz channels. This new standard should only be implemented if a new access point is added so that the WLAN requirements can be met along with the delivery of wireless customers' Ethernet speeds.

*(v) IEEE 802.11p*

IEEE802.11p amendment is designed to support Wireless Access in Vehicular Environments (WAVE). Vehicle-to-Everything (V2X) based applications used this standard in 5.9 GHz band to communicate with immediate surroundings like other vehicles termed as vehicle-to-vehicle (V2V), roadside infrastructure termed as vehicle-to-infrastructure (V2I) etc. A global standard is very essential so that smart and autonomous vehicles made by different manufacturers can interface with each other as well as other surrounding devices. The need of a fast and stable standard for V2X communication is of utmost important as any communication failure can lead to an accident and other undesirable situations [71].

*(vi) IEEE 802.11ac*

The IEEE 802.11ac Task Group (TGac) has developed this amendment with the goal to reach a aggregate network throughputs of at least 1 Gbps under 6 GHz bands that does not include the 2.4 GHz band. For a single user, this standard assumes a maximum MAC thoughput of at least 500 Mbps, and 1000 Mbps for for multiple users. In comparison to the all previous standards, the 802.11ac standard improves the throughput of complete network as well as performance of individual link. Due to this significant increased performace maintained by 802.11ac, this amendement is also reffered to as very high throughput (VHT) [72].

## 7. CONCLUSION AND FUTURE SCOPE

Efficient and effective MAC protocols for WMN have received immense attention from researchers and academia. Significant trials have been discussed in this manuscript that faces MAC protocol design and development, then after the design requirements for facilitating communication in IEEE 802.11 networks, such as IEEE 802.11a, 802.11b, 802.11g, 802.11n, 802.11ac, and 802.11p are examined. IEEE 802.11ac, which works at 5 GHz, is the newest

framework for achieving Gigabits per second capacity transfer rates and has better performance and fewer interferences. It has spectrum widths of 20, 40, 80, and 160 MHz.

The future scope of IEEE 802.11 is the IEEE WirelessHUMAN (Wireless High-speed Unlicensed MAN). IEEE WirelessHUMAN project is working on the standards for fixed wireless access in the license-free frequency. The MAC layer of WirelessHUMAN will be modified from IEEE 802.16, and the physical layer would be based on IEEE 802.11a or related specifications' OFDM mechanism. The key characteristics of the two WLAN standards, IEEE 802.11a and HiperLAN/2, are combined and can be applied to Wireless HUMAN systems.

Although, the paper emphasized the theoretical approach in the comparative study of different standards, in future, we will make use of simulation software for an empirical comprehensive performance analysis of more recent protocols on various quality of service parameters. Further, the security issues and challenges related to WMN would be surveyed and evaluated to recommend possible solutions to handle dissimilar attack vectors in the future.

## APPENDIX: LIST OF ACRONYMS

| Acronyms | Full Name |
|---|---|
| WMN | Wireless Mesh Network |
| MAC | Media Access Control |
| MANET | Mobile Ad-hoc Network |
| GSM | Global System for Mobile Communications |
| WLAN | Wireless Local Area Network |
| APs | Access Points |
| CSPAN | Cable Satellite Public Affairs Network |
| CSMA/CA | Carrier Sense Multiple Access with Collision Avoidance |
| QoS | Quality of Service |
| PHY | Physical Layer |
| MPT/MPR | Multi-Packet Transmission/Reception |
| MPC | Concurrent Communication Message-Passing Concurrency |
| MU-MIMO | Multi-User Multiple Input Multiple Output |
| MS | Mobile Station |
| BS | Base Station |
| RS | Relay Station |
| SA | Security Association |
| PCF | Point Coordination Function |
| DCF | Distributed Coordination Function |
| HCF | Hybrid Connectivity Feature |
| EDCA | Enhanced Distributed Channel Access |
| HCCA | HCF Controlled Access |
| Vo-WLAN | Voice over Wireless LAN |
| FD-MAC | Flow Driven Media Access Control |
| PDR | Packet Delivery Ratio |
| MACA-P | Medium Access via Collision Avoidance with Enhanced Parallelism |
| MRU | Multi-Radio Unification |
| CDMA | Code Division Multiple Access |
| AODV | Ad-hoc On-demand Distance Vector Routing |
| DSR | Dynamic Source Routing |
| OLSR | Optimized Link-State Routing Protocol |
| GRP | Geographic Routing Protocol |
| EDCF | Enhanced Distributed Coordination Function |
| OFDM | Orthogonal Frequency Division Multiplexing |
| FHSS | Frequency-Hopping Spread Spectrum |
| DSSS | Direct Sequence Spread Spectrum |
| ERP | Extended Rate PHY |
| ERP-DSSS-CCK | Extended Rate Physical Direct Sequence Spread Spectrum Complementary Code Keying |
| ERP-OFDM | Extended Rate Physical Orthogonal Frequency Division Multiplexing |
| PBCC | Packet Binary Convolution Code |
| DSSS-OFDM | Direct Sequence Spread Spectrum Orthogonal Frequency Division Multiplexing |
| DSRC | Dedicated Short-Range Communications |
| CCA | Clear Channel Assessment |
| FHSS | Frequency Hopping Spread Spectrum |

## DECLARATIONS

**Authors' contributions:** Conceptualization: Ankita Sing, Sudhakar Sing; Shiv Prakash; Methodology: Ankita Sing, Sudhakar Sing, Shiv Prakash; Formal analysis and investigation: Ankita Singh, Sudhakar Singh, Shiv Prakash; Writing - original draft preparation: Ankita Singh, Sudhakar Singh; Writing - review and editing: Sudhakar Singh, Shiv Prakash; Resources: Ankita Singh, Sudhakar Singh, Shiv Prakash; Supervision: Sudhakar Singh, Shiv Prakash.

**Conflicts of interest:** The authors declare that they have no conflict of interest in this paper.

**Funding:** No funding was received for conducting this study.

**Availability of data and material:** Not applicable.

**Code availability:** Not applicable.